\documentclass[aps,prb,showpacs,twocolumn,floatfix]{revtex4}

\newcommand{\br}{{\bf r}} \newcommand{\bs}{{\bf s}}
\newcommand{\bq}{{\bf q}} \newcommand{\ik}{\kappa^{-1}}
\newcommand{\beq}{\begin{equation}} \newcommand{\eeq}{\end{equation}}
\newcommand{\hp}{\Phi}

\newcommand{\figwidth}{1 \columnwidth} 
\newcommand{\dfigwidth}{2 \columnwidth} 
\usepackage{amssymb}
\usepackage{graphicx}  
\usepackage{dcolumn} 
\usepackage{bm}

\begin{document}

\title{Columnar versus smectic order in systems of charged colloidal rods}
\author{H. H. Wensink } \email{wensink@thphy.uni-duesseldorf.de}
\affiliation{Institut f\"{u}r Theoretische Physik,
Heinrich-Heine-Universit\"{a}t D\"{u}sseldorf,
Universit\"{a}tsstra{\ss}e 1, D-40225, D\"{u}sseldorf, Germany}
\pacs{64.70.Md,82.70.Dd,61.20.Ja} \date{\today}

\begin{abstract} We study the stability of inhomogeneous liquid
crystalline states in  systems of monodisperse, stiff, charged rods.  
By means of a  bifurcation analysis
applied to the Onsager free energy for charged  rods in strongly
nematic states,
we investigate nematic-smectic and nematic-columnar instabilities as a function of
the Debye screening length $\ik$. While the nematic-smectic transition
clearly  pre-emts the nematic-columnar one in the regime of strong
screening (i.e. small $\ik$) a marked stability of  hexagonal
columnar order  is observed at larger screening lengths.  The
theoretical results are substantiated by Brownian  dynamics computer
simulation results based on the Yukawa-site model. Our findings
connect  to experiments on  tobacco mosaic virus rods in
particular but might be relevant for soft  rod-like mesogens in strong
external directional fields in general. \end{abstract}

\maketitle

\section{Introduction} 

Over the past decades, much research effort has been devoted to
understanding the liquid crystal phase behavior of hard
non-spherical colloidal particles, particularly in density functional theory and computer simulations. The
theoretical approach to lyotropic liquid crystal formation has been initiated a long time ago by Onsager
\cite{onsager} with his classic paper on the isotropic-nematic
transition of infinitely thin rods. This theory shows that repulsive
interactions alone can lead to long-range orientational (nematic) order,
disproving the notion that attractive interactions are a prerequisite. The full phase
diagram for hard spherocylinders with variable aspect ratio has only fairly recently been mapped out completely by
means of extensive computer simulations
\cite{mcgrother,Bolhuisintracing}. A notable feature is that
 nematic order is only stable for sufficiently large aspect ratios
while isotropic systems of short rods tend to
freeze directly into a crystalline state with three-dimensional
long-range positional order. As to the nematic state, additional
high-density phase
transitions may occur involving partial freezing into liquid crystals with
long-range spatial inhomogeneities in either one dimension, giving a smectic A (SmA) phase, or two dimensions leading
to columnar (C) order. The latter also plays a dominant role in the
high-density phase behavior of plate-like colloids where a  stable hexagonal columnar
 was found in simulations \cite{frenkellc,veerman}
and experimental studies of model clay suspensions  \cite{Brown99,vanderkooijcolumnar}.

Complicating issues, often important in interpreting experimental results,
include the effect of rod flexibility \cite{khokhlov81,odijkoverview} and size polydispersity \cite{onsager}.
Flexibility plays an important role in systems of e.g. stiff polymers and
linear micelles and its generic effect is a significant depression of
nematic order compared to rigid particles\cite{Vroege92}. Systems of  self-assembling worm-like micelles
may, under certain circumstances, not show  nematic order at all. Instead 
 a phase transition
from an isotropic fluid to a hexagonal columnar phase
takes place \cite{schoot-hex, taylor-self}.  The stability of the columnar state is also strongly stimulated by 
the inherent length polydispersity which may be of the annealed form,
where the size distribution depends upon density such as in micellar systems,
or of the quenched type like for  many colloidal systems where the
size distribution is fixed by the synthesis procedure. A simple
packing argument suffices to understand that rods with variable length
do not easily fit into layers pertaining to smectic order and
therefore prefer columnar order. The crossover
to columnar order has  been observed explicitly in binary mixtures of hard rods with two different lengths \cite{stroobants-bi,cuichen}.

Another important factor in the phase stability of rodlike mesogens
is the influence of the soft repulsive interactions.
 In the experimental situation, interactions between colloidal
model rods with chemically modified surfaces
 can never be rendered truly hard \cite{bruggen-ach,bruggen-shell} 
and  many important biomacromolecular systems such as tobacco mosaic 
virus (TMV) and {\it fd} virus rods \cite{fraden-tmv-baus}, or mineral
systems
 \cite{davidson-overview} like goethite \cite{davidsongoethiet} 
and vanadium pentoxide ($\text{V}_{2}\text{O}_{5}$) \cite{Zocher} rods are
stabilized by  electrostatic particle repulsions. 
The influence of these interactions on the
isotropic-nematic transition had already
been addressed by Onsager in his original paper. Later on, the effect of
electrostatic `twist', which disfavors parallel rod configurations and
hence destabilizes nematic order, has been
worked out in more detail in a study by Stroobants {\it et al.}
\cite{stroobantslading}. 

Much more challenging however is the question how the electrostatic
interactions affect the
stability of the inhomogeneous liquid crystal phases, in particular the
smectic phase, formed at high densities.  Theoretical attempts in this direction
 have been undertaken in Ref. \onlinecite{kramerherzfeld-pre} and \onlinecite{graftmv}.  Based on
a model which combines the concept of an effective rod thickness
(to account for the extent of the electric double layer) with a simple cell description to treat the inhomogeneous
states, Kramer and Herzfeld \cite{kramerherzfeld-jcp,kramerherzfeld-pre}  were able to construct the phase diagram of parallel,
charged  rods. Their  main conclusion, a generic charge-induced stabilization of the
inhomogeneous liquid crystal states also shows up in
the study of Graf and L\"{o}wen \cite{graftmv} based on an elaborate density
functional theory.

In this paper we present a similar study starting from the
 Onsager theory for freely rotating rods. Rather than attempting
to construct the full phase diagram we will merely focus on locating  symmetry-breaking instabilities of the
nematic state towards the inhomogeneous phases. The main candidates are
 the smectic and the hexagonal columnar phases. The study of these type of instabilities 
has been pioneered a few decades ago by Mulder
\cite{mulderbifur, mulderparns} and  the preferential stability of smectic over columnar
order in systems of freely rotating hard rods, first recognized in simulations \cite{Frenkel88},
could also be established in theory \cite{poni1988}.

Here, we will elaborate on the  Onsager-type
approaches  and extend  the  calculations towards charged systems in two steps. First,
systems of  parallel charged rods will be considered by employing a
 straightforward site model to account for the electrostatic end cap effects. Next, the  restriction of parallel confinement
will be removed and the influence of rotations and electrostatic twist
shall be explicitly taken into account by considering the pair potential for infinitely stretched
line charges.  Although the second virial theory is not quantitatively valid
for our case (the weight of parallel 
rod configurations  which drive the translational symmetry-breaking instabilities 
 necessitates inclusion of higher virial coefficients into the free energy),
we expect the theory to give a reliable qualitative sketch of the
competitive stability of smectic and columnar order for soft rods.

In contrast to the previous theoretical studies of
Refs. \onlinecite{kramerherzfeld-pre} and \onlinecite{graftmv}, our calculations reveal a distinct crossover
from smectic to columnar order upon decreasing ionic strength. 
These results may be helpful in interpreting experimental results on TMV
rods at low ionic strength. Most importantly, we show that columnar stability in rod systems
need not be induced  by length polydispersity but may be brought
about by the soft electrostatic interactions only. Moreover, it is shown
that the region of manifestation of columnar order can be broadened
considerably if the isotropic phase is suppressed. This can be
achieved by  applying a strong external directional field. Preliminary Brownian dynamics
simulations based on a Yukawa-site model are also carried out and the results
point to a marked stability of columnar textures at low screening conditions for asymptotically
aligned rods, in agreement with theory.

This paper is organized as follows. In Sec. II the general  Onsager
functional is introduced and the bifurcation analysis  is outlined in
Sec III. The subsequent sections deal with the explicit calculation of the
appropriate Mayer kernels which form the necessary ingredients of the
analysis. This is done first for parallel charged rods in Sec. IV and
then for freely rotating hard rods (Sec. V) and charged ones (Sec. VI).
Sec. VII  is devoted to the  Brownian dynamics simulations and
all results will be  combined  and discussed in Sec. VIII. Finally, some
concluding remarks  will be formulated in Sec. IX.

\section{Onsager functional} The starting point of our analysis is to
construct a general free energy functional for an inhomogeneous system
of $N$ freely rotating rods with  arbitrary mutual interactions in a
volume $V$. The  simplest approach is to take Onsager's well-known
result \cite{onsager} for a virial expansion of the Helmholtz  free energy $F$ truncated after
the first non-trivial  term, which we may recast into the following
functional 
\begin{eqnarray}
\beta F[\rho] &=& \int \rho(\bs) \{ \ln {\mathcal V}
\rho(\bs) - 1 \} d \bs   \nonumber \\
&&  - \frac{1}{2} \int d \bs  \rho(\bs) \int d \bs ^{\prime} \rho(\bs
^\prime)  \Phi(\bs,\bs ^\prime)  \label{free}
\end{eqnarray} 
with $\beta^{-1} = k_{B}T $ the thermal energy ($k_{B}$ is
Boltzmann's constant and $T$ temperature) and $\mathcal V$ the thermal
volume of a rod.  The rod  density distribution $\rho$ is a function
of the generalized  coordinates $\bs = \{ \br,\Omega \}$, indicating
position and solid angle,  respectively and is normalized according to
$\int d \bs \rho(\bs) = N/V$. The first term in Eq. (\ref{free})
representing the ideal mixing free  energy is  exact while the second
contribution contains the Mayer function $\Phi(\bs,\bs ^\prime)$ to
account for the rod interactions on the approximate pairwise
level. The Mayer function is related to the direct rod pair
interaction energy $w$ via
\beq  
\Phi(\Delta \br;\Omega,\Omega ^\prime) = \exp [
-\beta w(\Delta \br;\Omega, \Omega ^\prime)] - 1 \label{mayer}
\eeq
 which depends  on the centre-of-mass difference
vector $\Delta \br = \br ^{\prime} - \br$ of a pair of rods and their
orientations. For hard rods  $\beta w $ is infinitely large if the particles
overlap and zero otherwise. If the system is also homogeneous, the
density distribution is independent of  position and can be factorized
into $\rho = f(\Omega) N/V$, where $f(\Omega)$ is a
distribution of orientations obeying $\int d \Omega f(\Omega)=1$. Spatial integration of
Eq. (\ref{mayer}) then gives the relation
\begin{equation} \int  d \Delta \br \Phi(\Delta \br;\Omega, \Omega
  ^{\prime}) = -v_{\text{excl}}(\Omega,\Omega ^{\prime}) \label{vex}
\end{equation} with $v_{\text{excl}} = 2L^{2}D |\sin \gamma
(\Omega,\Omega^{\prime})|$, the excluded volume of two thin
hard rods with length $L$ and diameter  $D$  (such that $L/D \gg 1$) at interrod angle $\gamma$. In
general, Eq. (\ref{vex}) forms the basis of the classic Onsager-type
theories for homogeneous systems of hard  anisometric particles
\cite{wensinkthesis}. For any given density $N/V$ the thermodynamics
of the system is fully contained in the orientation  distribution
function  $f(\Omega)$ (ODF). Its equilibrium form can be uniquely
obtained for a given density by requiring the free energy to be
minimal. Formally minimizing  the free energy with respect to the ODF
leads to the following general stationarity equation for
homogeneous systems \beq f( \Omega ) = {\cal N}^{-1 }\exp \left [ \rho
\int d \Omega ^{\prime}  f( \Omega ^{\prime}) \int d \Delta \br \Phi ( \Delta \br ; \Omega, \Omega
^{\prime})  \right ] \label{stat} \eeq
with ${\cal N} = \int d \Omega ^{\prime} \exp[\cdots] $ to ensure
normalization.  Upon increasing density, a change of shape of
$f(\Omega)$ from a constant to a peaked distribution signifies an
orientational symmetry-breaking transition of the system from an
isotropic state (with random rod orientations) towards e.g. a nematic one
where the rods point in globally the same direction, defined as the
nematic director.

At higher densities, additional phase transitions from the  nematic
phase towards states with broken translational symmetry such as
smectic or columnar phases can be expected. A convenient way to
approximately locate phase transitions of this kind is to apply a bifurcation
analysis \cite{kayser,mulderbifur} to the generalized Onsager free
energy functional.

\section{Bifurcation analysis}

In a smectic or columnar liquid crystal, the density is no longer
constant throughout space but shows periodic spatial modulations in
one dimension along the nematic director (smectic order)
 or in a two-dimensional Bravais lattice (hexagonal, cubic,
et cetera) perpendicular to the director (columnar order). 
We may thus  propose the following Fourier
expansion for the density distribution:
 \beq 
\rho( \br,\Omega) = \rho \sum _{k} \sum _{l \ge 0} a_{l}(\Omega ) \exp
[ i l \bq _{k} \cdot
\br ] \label{rhofourierformal} 
\eeq 
in terms of the orientation dependent amplitudes $a_{l}(\Omega)$  quantifying positional order along the various
Fourier modes related to the wave number $q=2  \pi/\lambda$, with
$\lambda$ the typical spacing pertaining to the density
modulations. While for smectic order a single wavevector ($k=1$)
suffices, implementing columnar (or crystalline) order requires a
superposition of different
 wave vectors $ \{ \bq _{1}, \bq _{2},\cdots \} $ to
reproduce the desired lattice.
 We will come back to this later. Identifying $a_{0}(\Omega) = \rho
 f(\Omega )$ for the homogeneous system, we may simplify the above expansion
as follows (omitting the $k$-summation for clarity):
 \beq 
\rho( \br,\Omega) = \rho
f ( \Omega ) \left  [ 1 + \sum _{l \ge 1} a_{l} \cos ( l \bq \cdot
\br ) \right  ] \label{rhofourier} 
\eeq 
where we used that $\bq = -\bq$. The ODF $f(\Omega)$ is now taken  to be that  of the nematic reference
state for {\em all} modes $l$. This  means
that any coupling between fluctuations in the spatial density  and the
orientations is neglected. A fully consistent  calculation for the nematic-smectic
bifurcation \cite{roijtransverse} shows that the position-orientation
coupling is marginally small for strongly aligned,  slender rods  on
which we focus in this study.

Inserting Eq. (\ref{rhofourier}), truncated after $l=1$, into the free energy and expanding
the free energy of the new inhomogeneous (I) state with respect to the
homogeneous nematic (N) one $\delta F = F_{I} -F_{N}$ up to quadratic
order gives the free energy curvature 
\begin{eqnarray} 
\delta^{2} \beta F &=&  a_{1}^{2} \rho^{2} \int  d \Omega f(\Omega)
\int d \Omega ^{\prime} f(\Omega ^{\prime}) \nonumber \\
&& \times  \left ( \frac{\delta (\Omega, \Omega^{\prime}) }{\rho f(\Omega)} - {\hat
\Phi}(\bq ; \Omega, \Omega^{\prime}) \right ) 
\end{eqnarray}
The nematic state becomes locally unstable if the
curvature vanishes, $\delta ^{2} \beta F = 0$. The above expression
then simplifies to the {\em bifurcation} condition 
\beq 
\rho  \int d \Omega  f(\Omega) \int  d\Omega ^{\prime} f(\Omega ^{\prime}) {\hat \Phi}(\bq ; \Omega,
\Omega^{\prime})  = 1 \label{bif} 
\eeq 
where the hat denotes a cosine transform of the Mayer function: 
\beq 
{\hat
\Phi}(\bq ; \Omega, \Omega^{\prime}) = \int d \Delta \br \Phi (\Delta \br; \Omega,
\Omega^{\prime}) \cos (\bq \cdot \Delta \br )  
\eeq 
which is the key input of the analysis. The bifurcation density  is defined
as the smallest non-trivial physical solution $\rho^{\ast}$ of Eq. (\ref{bif}) with
associated wave vector $\bq ^{\ast}$.  Once the instability has been
located, additional information about the thermodynamic stability and
the order of the phase transition can be inferred from a
parametric expansion about the bifurcation point, as outlined in
Ref. \onlinecite{mulderparns}. In Appendix B we shall reproduce the analysis
and  derive a general expression for the  Landau free energy
of the inhomogeneous state with respect to the homogeneous nematic
system. This result will then be used to verify the thermodynamic
stability of the smectic and columnar states. In the next sections we
will derive expressions for ${\hat {\Phi }}(\bq)$ for the  nematic
phase  focussing first on {\it parallel} charged rods. The theory is
then extended to include the effect of rotations.

\section{Parallel charged rods: hard-core Yukawa site model}

To calculate the  Mayer kernel for parallel {\it charged} rods we have
to find a suitable route to include the influence of electrostatic
{\it end effects}.  These end effects are intricately difficult to quantify
\cite{odijkendeffects} but are nevertheless  essential in our description
since the formation of  stable smectic/columnar liquid crystalline
structures at high densities is driven primarily by correlations
between the end cap of one rod and the main (cylindrical) manifold of
the other as  embodied in the ${\cal O}(LD^{2})$ contributions to the
Mayer kernel. To make headway  we consider a  model in which a rod with
finite length $L$ is composed of an array of $n \gg 1$ spherical beads
with diameter $D$ placed at equidistant intervals.  Each bead $i$ ($ 1 \leq
i \leq n$) from one
rod interacts with  bead $j$ from the other via a hard-core Yukawa
(HCY) potential. The total rod potential depends only on the
 centre-of-mass
distance ${\bf s}$ (expressed in units $D$) and is given by
\begin{equation} \beta w _{el}^{HCY} ({\bf s}) =  \left \{
\begin{tabular}{lll} $ \epsilon \sum _{i} \sum _{j} \frac{\exp \left [
- \kappa D \left ( s_{ij} - 1 \right ) \right ] }{s_{ij}}$ & if all &
$ s_{ij} \geq 1$ \\  &  &  \\  $\infty $ & if any & $ s_{ij} < 1$
\end{tabular} \right.  \label{hcyuk}
\end{equation} 
with $\kappa$ the Debye screening constant and
$\epsilon$ a dimensionless  contact potential $\epsilon = (Z/n)^{2}(
\lambda _{B}/D)/(1+\kappa D/2)^{2}$ depending on the total rod charge
$Z$ (in units of the elementary charge $e$) and the Bjerrum length
$\lambda_{B} = \beta e^{2}/\varepsilon _{0} \varepsilon _{r}$ which is
determined by the temperature and the dielectric constant $\varepsilon_{r}$ of the
solvent.  The quantity $s_{ij}$ represent the distance (in units $D$)
between site $i$ from one rod and $j$ from the other. It is
convenient to introduce cylindrical coordinates, so that $ {\bf s} =\{
s_{\perp}\cos \omega, s_{\perp} \sin  \omega, s_{\parallel} \}$, with
$s_{\perp}$ expressed in units $D$ and $s_{\parallel}$ in units $L$.
The site-site distance $s_{ij}$ is then given by 
\beq 
s_{ij} = \sqrt{s_{\perp}^{2} + (L/D)^2 [ s_{\parallel} + d(j-i) ]^2},
\hspace{0.5cm} i,j=1,2,\cdots n 
\eeq 
Here, $d = (L/D)/(\sqrt{n+1}\sqrt{n-1})$ is the dimensionless spacing between two
neighboring sites on a rod, chosen such that the quadrupolar moment \cite{kirch-bd}
of the rod matches that of a homogeneous line charge of length $L$.

The total transformed Mayer function for the HCY-site model is then
given by: 
\beq 
\hat{\Phi}_{HCY}({\bf q}) = LD^{2} \int d {\bf s}
\Phi _{HCY} ( {\bf s}) \cos (D \bq \cdot {\bf s } ) 
\eeq 
The generalized wave vector is given by $\bq = \{q_C,0,q_{S} \}$ in terms
of smectic (S) and columnar (C) components. Hence, we can write  $D  \bq \cdot
{\bf s} = Q_{C} s_{\perp} \cos \omega + Q_{S} s_{\parallel} $, with
the dimensionless wave numbers  
\beq
Q_{S} = 2\pi \left( \frac{L}{\lambda _{S}} \right ), \hspace{1cm} Q_{C} = 2\pi
\left ( \frac{D}{\lambda _{C}} \right ) \label{dwn}
\eeq
related to the lattice spacing $\lambda$.

Integrating over the angle $\omega $ then yields for the {\it
electrostatic contribution} (denoted by ``Y'') to the Mayer kernel:
\begin{widetext}
\begin{eqnarray} \hat{\Phi}_{Y}(Q_{C},Q_{S}) & = & 2 \pi LD^{2} \int
_{0} ^{\infty} 2 d s_{\parallel} \int _{1} ^{\infty} s_{\perp}
ds_{\perp} \Phi_{HCY} (s_{\perp},s_{\parallel}) J_{0}(Q_{c}s_{\perp})
\cos (Q_{S}s_{\parallel})  \nonumber \\ && +  2 \pi LD^{2} \int _{1}
^{\infty} 2 d s_{\parallel} \int _{0} ^{1} s_{\perp} ds_{\perp}
\Phi_{HCY} (s_{\perp},s_{\parallel})  J_{0}(Q_{c}s_{\perp}) \cos
(Q_{S}s_{\parallel}) \label{yuk}
\end{eqnarray}
\end{widetext}
 where the spatial integration has been performed over
the region outside the cylindrical manifold determined by the
hard-core excluded volume of two parallel cylinders.  The
complementary  integral over the inner region where the cylindrical
cores overlap yields the hard-core contribution (denoted by ``H'')
\begin{eqnarray} \hat{\Phi}_{H}(Q_{C},Q_{S})  &=&  2 \pi LD^{2} \int
_{0} ^{1} 2 d s_{\parallel} \int _{0} ^{1} s_{\perp} ds_{\perp}
\Phi_{HCY} \nonumber \\
&& \times  J_{0}(Q_{c}s_{\perp}) \cos Q_{S}s_{\parallel }
\nonumber \\
 &=&  - 4 \pi L D^{2} j_{0}(Q_{S}) \frac{J_{1}(Q_{c})}{Q_{c}}
\label{parhard}
\end{eqnarray} with $j_{0}(x)=\sin x/x$ a spherical Bessel function
and $J_{k}$ a standard one. In addition we used that  $\Phi_{HCY} =-1$
for overlapping hard bodies. The result is
identical to that of Ref. \onlinecite{mulderparns}.  We remark that the excluded
volume between two spherical end caps $\sim {\cal O}(D)^3$ is not resolved exactly by the
integrations of Eqs. (\ref{yuk}) and (\ref{parhard}) and the Yukawa
contribution in Eq. (\ref{yuk}) is therefore  expected to be reliable up to ${\cal
O}(LD^2)$, which suffices for slender rods. The integrations in
Eq. (\ref{yuk}) can be carried out numerically by imposing appropriate
cutoff distances for $s_{\perp}$ and $s_{\parallel}$. The bifurcation
condition Eq. (\ref{bif}) is obtained by substituting  $\hat {\Phi} =
\hat{\Phi} _{H} + \hat{\Phi}_{Y}$  and $f(\Omega)=\delta(\Omega)$.

\vspace{0.3cm}

\section{Freely rotating hard rods}

Before embarking on our calculation of $\hat{\Phi}$ for freely
rotating charged rods, we will first discuss the effect of rotations
on the translational symmetry-breaking bifurcations in systems of
hard rods. 
A closed expression for $\hat{\Phi}(\bq)$ for freely rotating,
{\it hard} spherocylinders has been derived by van Roij
\cite{vroijthesis}:
\begin{widetext}
\begin{eqnarray} {\hat \Phi}_{H}(\bq; \Omega_1, \Omega_2  ) & =&   -  2
L^{2} D | \sin \gamma | j_{0} (D \bq \cdot {\bf  \hat{v} } ) j_{0}
\left ( \frac{L}{2} \bq \cdot  {\bf  \hat{w}_{1} } \right ) j_{0}
\left ( \frac{L}{2} \bq \cdot { \bf \hat{w }_{2} } \right )  \nonumber
\\  && -   L D^{2}  j_{0} \left ( \frac{L}{2} \bq \cdot {\bf
\hat{w}_{1} } \right ) \left [2 \pi \cos \left ( \frac{L}{2}  \bq
\cdot {\bf  \hat{w}_{2} } \right   ) \frac{J_{1}(M_{1})}{M_{1}} + 2
\sin \left ( \frac{L}{2}  \bq \cdot {\bf  \hat{w}_{2} } \right )
W(M_{1},  \phi _{1} ) \right ] \nonumber \\ && -   L D^{2}  j_{0}
\left ( \frac{L}{2} \bq \cdot {\bf  \hat{w}_{2} } \right ) \left [2
\pi \cos \left ( \frac{L}{2}  \bq \cdot {\bf  \hat{w}_{1} } \right   )
\frac{J_{1}(M_{2})}{M_{2}} - 2 \sin \left ( \frac{L}{2}  \bq \cdot
{\bf  \hat{w}_{1} } \right ) W(M_{2},  \phi _{2} ) \right ]  \label{phitot}
\end{eqnarray} 
\end{widetext}
which contains all contributions up to ${ \cal O}
(LD^2)$. Correction terms of ${\cal O}(D^3)$, for which no closed
analytical expressions are  available, can be safely neglected for
sufficiently anisometric rods $L/D \gg 1$. The inner products contain
the rod orientation unit vectors $ {\bf   \hat{w}_{1}}$ and ${\bf
\hat{w}_{2} }$ with associated unit vector ${\bf  \hat{v}} 
= {\bf  \hat{w}_{1} } \times {\bf  \hat{w}_{2} } / \sin \gamma$.  The function $W$
is defined as 
\beq 
W(M, \phi) = \int_{-\pi/2}^{\pi/2} d \alpha j_{1}(M \cos (\alpha - \phi)) 
\eeq
with $j_{1}(x) = x^{-2} \sin x - x^{-1} \cos x$ a
spherical Bessel function.

Defining a third unit basis vector $ {\bf \hat{u} } _{i}
= { \bf \hat{w} } _{i} \times { \bf \hat{v} } $ the quantity $ M_{i}
$ and the angle $ \phi _{i}$ are given by
\begin{eqnarray} 
M_{i} &=& Dq \sqrt{ ({\bf \hat{q}} \cdot {\bf
\hat{v}})^{2}   +   ({\bf \hat{q}} \cdot {\bf \hat{u}}_{i})^{2} },
\hspace{0.5cm} i=1,2
\nonumber \\  \phi _{i} &=& \arccos \left (  \frac{ {\bf \hat{q}} \cdot {\bf
\hat{u}} _{i} }   { \sqrt{ ({\bf \hat{q}} \cdot {\bf \hat{v}})^{2}   +
({\bf \hat{q}} \cdot {\bf \hat{u}}_{i})^{2} } }    \right )
\label{anglephi}
\end{eqnarray} 
The result for perfectly aligned rods is easily
recovered by replacing the orthonormal set of basis vectors  $\{ {\bf
\hat{u}}_{i}, {\bf \hat{v}}, {\bf \hat{w}}_{i} \}$ by the set of
Cartesian unit vectors.  We may thus identify  ${\bf \hat{u}}_{1} =
{\bf \hat{u}}_{2} = \{ 1,0,0 \} $, ${\bf \hat{v}} = \{ 0,1,0 \} $ and
${\bf \hat{w}}_{1} = {\bf \hat{w}}_{2}=\{ 0,0,1 \} $.  With  help of
the identity $j_{0}(x)=\cos(x/2) j_{0}(x/2)$ Eq. (\ref{phitot}) then
immediately gives back Eq. (\ref{parhard}). 

Let us now perform an
asymptotic analysis of the Mayer kernel, valid for strongly nematic
systems  where the average deviation of the rod vectors from the nematic director
is small. 

\subsection{Smectic symmetry}

For the smectic case, density modulations occur along the  nematic
director ${\bf \hat{n}} $  fixed at  ${\bf \hat{n} } =\{ 0,0,1\}$ and
hence  ${\bf q} = q_{S}\{ 0,0,1 \}$. The orientation of rod $i$ on
the unit sphere can be parametrized  in terms of a polar angle
$\theta_{i}$ and an azimuthal one $\varphi_{i}$, so that  $ {\bf
\hat{w}}_{i} = \{\sin \theta_{i} \sin \varphi_{i}, \sin \theta_{i}
\cos \varphi_{i}, \cos \theta_{i} \}$.  Expanding all relevant inner
products for small polar angles $\theta_{i}$ up to leading order gives:
\begin{eqnarray} \frac{L}{2}{\bf q} \cdot {\bf \hat{w}}_{i} &\sim&
\frac{1}{2} Q_{S}  \nonumber \\ D{\bf q} \cdot {\bf \hat{v}} &\sim&
Q_{S} \frac{D}{L} \left ( \frac{ \theta_{1}\theta _{2} } { \sqrt{
\theta ^2_{1} + \theta ^2_{2} - 2 \theta _{1} \theta _{2} \cos \Delta
\varphi} }   \right ) \nonumber \\ M_{i} &\sim&  Q_{S} \frac{D}{L}
\theta_{i}
\end{eqnarray} with $\Delta \varphi = \varphi_{2} - \varphi_{1}$. For
large aspect ratios ($L/D \gg 1 $) all contributions of ${\cal O}\left
[(D/L) \theta _{i} \right ]$ become marginally small so that the
following limiting values can be deduced : $M_{i} \rightarrow 0$ and
$(D{\bf \hat{q}} \cdot {\bf \hat{v}}) \rightarrow 0$. Consequently,
$J_{1}(M)/M \rightarrow 1/2$ and $W(M,\phi) \rightarrow 0$.  Using
this  in Eq. (\ref{phitot}) together with the identity
$j_{0}(x)=\cos(x/2) j_{0}(x/2)$ gives the following asymptotic
expression for the transformed Mayer function for the smectic
symmetry: 
\begin{eqnarray}
{\hat \Phi}_{H}(\bq; \Omega _1, \Omega _2  ) &=&  -  2
L^{2} D | \sin \gamma (\Omega _1, \Omega _2 )|  j_{0} ^{2} ( Q_{S}/2
)  -\nonumber \\
&&  2 \pi  L D^{2}  j_{0} (Q_{S}) \label{smecor} 
\end{eqnarray}
The latter contribution is simply the result for {\it parallel} rods, given by
Eq. (\ref{parhard}) taking the limit $Q_c \rightarrow 0$, while the
first is the leading order correction term arising from the rotations.

The bifurcation condition is obtained by performing an orientational
average according to Eq. (\ref{bif}). Following Odijk
\cite{OdijkLekkerkerker}, we introduce the Gaussian trial ODF
 \begin{equation} f_{G}(\theta ) \cong \frac{\alpha}{4 \pi}  \left\{
\begin{tabular}{lll} $ \exp \left [-
\frac{1}{2}\alpha  \theta ^{2} \right ]$ & if & $ 0\leq \theta \leq
\frac{\pi }{2}$ \\  &  &  \\  $  \exp \left [-
\frac{1}{2} \alpha (\pi -\theta ))^{2} \right ]$ & if &
$\frac{\pi}{2} < \theta \leq \pi $%
\end{tabular} \right.  \label{0ODF}
\end{equation} 
which depends only on the polar angle in case of uniaxial nematic
order.  The variational parameter $\alpha$ is chosen such as to
minimize the nematic free energy. It must be much larger than unity because the normalization factor is only valid 
in this limit. After minimization $\alpha$ attains a quadratic density
dependence, $\alpha
= 4c^{2}/\pi$ where $c$ is a dimensionless rod concentration $c =
(\pi/4) \rho L^{2}D $ related to the {\it volume fraction}
$\eta  $ via $c= \eta L/D$. We may now use the following asymptotic
result \cite{onsager}: 
\beq
\langle  \langle |\sin \gamma | \rangle \rangle _{f_{G}}
 \sim  \left ( \frac{\pi}{ \alpha} \right )^{1/2} \label{avsin} 
\eeq
where the brackets denote an orientational
average. With this, the bifurcation equation Eq. (\ref{bif}) finally
becomes 
\beq -4 j_{0}^{2}(Q_{S}/2)  - 8 \eta j_{0}(Q_{S}) = 1 \label{smfinal} 
\eeq 
Note that the rotational  correction term is independent of the
density.

\subsection{Columnar symmetry}

Let us now turn to the columnar state  where  a two-dimensional
modulatory density pattern $\perp {\bf \hat{n}}$ is expected. In case
of a hexagonal Bravais lattice, a combination of the following
three unit wavevectors: 
 \beq 
\hat{\bq} _{1} = \left ( \begin{tabular}{c} $0$  \\   $1$  \\    $0 $ \end{tabular} \right ),
\hspace{0.2cm} 
\hat{\bq} _{2} = \left ( \begin{tabular}{c}  $ \sqrt{3}/2 $  \\  $ 1/2
$  \\  $0$  
\end{tabular} \right ),
\hspace{0.2cm}
\hat{\bq} _{3} = \left ( \begin{tabular}{c} $ -\sqrt{3}/2 $ \\  $ 1/2 $ \\ $ 0  $ \end{tabular} \right )
\eeq 
correctly reproduces the six-fold symmetry of the columnar density
modulations in the $xy$ plane. We may now repeat the asymptotic analysis to obtain the leading order expressions for
the inner products. For the ${\bf q}_{1}$-mode we obtain
\begin{eqnarray} \frac{L}{2}{\bf q }_{1} \cdot {\bf \hat{w}}_{i}
&\sim& \frac{1}{2} Q_{C} \frac{L}{D}  \theta_{i} \sin \varphi _{i}
\nonumber \\ D{\bf q }_{1} \cdot {\bf \hat{v}} &\sim& Q_{C} \left (
\frac{ - \theta_{1} \cos \varphi_{1} + \theta _{2} \cos \varphi _{2}
}{ \sqrt{ \theta ^2_{1} + \theta ^2_{2} - 2 \theta _{1} \theta _{2}
\cos \Delta \varphi}  } \right )\nonumber  \label{scalecol} \\ M_{i}
&\sim&   Q_{C}
\end{eqnarray} The expressions for the other columnar modes show
exactly the same scaling with respect to $\theta$ and $L/D$ and are not shown.  A quick
inspection reveals that the asymptotic analysis does not bring much
relief since i) $({\bf \hat{q}} \cdot {\bf \hat{v}}) \sim {\cal O}(1)$
and ii) the combination $(L/D) \theta_{i}$ appearing in the first line
also remains finite after having performed the orientational
average. It can be estimated with use of the Gaussian ODF
Eq. (\ref{0ODF}) and the resulting scaling relation reads $\langle
(L/D) |\theta | \rangle _{f_{G}} \sim \eta^{-1} \sqrt{\pi/2}
\sim {\cal O}(1)$.

Unlike for the smectic
case, the complicated nature of the orientation dependent terms
precludes an analytical calculation of their Gaussian
averages. Nevertheless, the integration over the angles $\theta _i$ and
$\varphi_i$  can be carried out  numerically without difficulty using standard Simpson
quadrature \cite{numrecipes}.
 The bifurcation condition Eq. (\ref{bif}) for the columnar symmetry is
given by a linear superposition of kernels 
\beq 
 \frac{\rho}{3}
\sum_{k=1}^{3} \left \langle \left \langle \hat{\Phi}_{H}({\bf q}_{k};
\theta _1, \varphi _1 ; \theta _2, \varphi _2 )    \right \rangle
\right \rangle _{f_{G}} = 1 \label{colbif} 
\eeq 
substituting the asymptotic forms Eq. (\ref{scalecol}) 
 in  Eq. (\ref{phitot}) and using the Gaussian ODF Eq. (\ref{0ODF}). 

\subsection{Bifurcation points}

The solutions of Eqs. (\ref{smfinal}) and  (\ref{colbif})
have been collected in Table I. To obtain more realistic volume fractions
a density rescaling according to  Parsons-Lee (PL) method\cite{parsons,Lee87,Lee89} has been
carried out. The approach, not treated here,  can be
implemented quite simply by rescaling the density according to $\rho
\rightarrow \rho g_{CS}(\eta)$, and equivalently   $\eta
\rightarrow \eta g_{CS}(\eta)$, where $g_{CS}(\eta) =
(1-(3/4)\eta)/(1-\eta)^2$ originates from the Carnahan-Starling
equation of state for a hard sphere fluid.  All volume
fractions mentioned throughout the rest of the paper are obtained from
this treatment.

It is clear that both N-S and N-C are destabilized due to the effect
of rotations. The decrease of the smectic layer spacing can be intuitively understood from  the fact
that the rod length projected onto the nematic director (the
`entropic' length) is smaller than the bare rod length due to the
orientational  fluctuations. Similarly, the increased columnar spacing 
can be attributed to an entropic rod diameter larger then the bare one.
 The N-S  pre-empting the N-C one  is in accordance with  simulations
\cite{Frenkel88,FrenkelJPC88} and experimental observations
\cite{fraden-tmv-baus}. In addition, the location of the nematic-smectic
transition predicted here compares very well with the value
$\eta_{S}^{\ast} \approx 0.418$ reported from simulations
\cite{Bolhuisintracing}.

\begin{table}
\begin{center}
\begin{tabular}{|l||l|l|l|l|} \hline
$$ & $\eta_{S}^{\ast}$ & $Q_{S} ^{\ast}$ & $ \eta_{C}^{\ast}$ & $ Q_{C} ^{\ast} $ \\ \hline

parallel rods  & 0.575 & 4.493 & 0.945 & 5.136 \\ \hline 

parallel rods (PL)  & 0.338 & 4.493 & 0.441 & 5.136 \\ \hline

rotating rods  & 0.792 & 4.858 & 1.540 & 4.424 \\ \hline

rotating rods (PL)  & 0.404 & 4.858 & 0.543 & 4.424 \\ \hline

\end{tabular} \vspace{0.5cm}
\caption{Overview of the nematic-smectic and nematic-columnar bifurcations for hard rods. (PL) refers to Parsons-Lee theory.}
\end{center}
\end{table}

\section{Freely rotating charged rods}

Looking back at the expressions derived for hard rods, in particular
Eq. (\ref{smecor}),  we may conclude
that the asymptotic Mayer kernel consist of two parts. First,  a `reference'  part for (near)-parallel
configurations associated with the ${\cal O}(LD^2)$ contributions in
Eq. (\ref{phitot})) and, second,  a part of ${\cal O}(L^2 D)$ which constitutes
a non-vanishing contribution in the limit of asymptotically strong alignment.  
In this section we will first  derive the leading
order ${\cal O}(L^2 D)$ correction due to electrostatic effects. 
These correction terms will then, together with  previous results,
be compiled into expressions for the {\it total}  Mayer kernel for
charged rods in near-parallel configurations.

Following Ref. \onlinecite{stroobantslading} we start with introducing the
following form for the electrostatic interaction between
two infinitely long charged rods at shortest distance $x$ and mutual
angle $\gamma$: 
\beq 
\beta w_{el} (x ; \Omega_{1}, \Omega_{2}) = A
\frac{  \exp [- \kappa (|{\bf x}|- D) ] } {| \sin \gamma (\Omega _{1},
\Omega_{2})| } \label{debyehuckel} 
\eeq 
If the rod charge is not too
high, the prefactor $A$ can be expressed in closed form within the
Debye-H\"{u}ckel approximation \beq A = \frac{8 \pi (\nu
\lambda_{B})^{2} \exp [-\kappa D]}{ (\kappa D)^{3} (\lambda _{B}/D)
K_{1}^{2}(\kappa D/2)} \label{aa} \eeq with $K_{1}(x)$ a modified
Bessel function. This expression stems  from the linearized
Poisson-Boltzmann equation for an infinitely long,  charged cylinder
with diameter $D$. The key quantities  here are the linear charge
density $\lambda _{B} \nu$ expressed in terms of unit charges per
Bjrerrum length and the ratio of the electric double layer thickness
and the rod diameter $1/\kappa D$.  The denominator in Eq. (\ref{debyehuckel})
embodies the electrostatic `twisting' effect whereby a rod pair is
energetically stimulated to adopt a perpendicular
configuration. Clearly,  for strictly parallel configurations, the expression
is ill-defined which means that a N-I bifurcation analysis cannot be
based solely on this potential. 

The electrostatic line charge contribution of
the cosine transform of the Mayer kernel can now be written as   
\begin{eqnarray}
{\hat \Phi}_{el}(\bq; \Omega_1, \Omega_2  ) &=&   \int \limits_{|{\bf x}| > D}
d |{\bf x}| \left \{ \exp \left [ -\beta w_{el} (|{\bf x}| ;
\Omega _1, \Omega _2) \right ] - 1  \right \} \nonumber \\ 
&& \times  \cos (\bq \cdot {\bf x})
\end{eqnarray}
If we neglect end effects, the shortest distance unit vector ${\bf
\hat{x}} $  is equal to $ {\bf \hat{v}}$ and the following parametrization ${\bf
x} = Dt {\bf \hat{v}}$, with $1<t< \infty$, can be applied. Rewriting
the integral gives
 \begin{eqnarray}
{\hat \Phi}_{el} &=&
 D \int _{1}^{\infty} dt \left \{  \exp \left [ -(A^{\prime}/|\sin
\gamma|) e ^{-\kappa D t} \right  ] -1 \right \} \nonumber \\ 
&& \times \cos (tD \bq \cdot {\bf
\hat{v}}) \label{phiel0} 
\end{eqnarray}
with $A^{\prime} = A \exp[\kappa D]$. 
To solve the integral it is advantageous to make the following
substitution $u(t) = (A^{\prime}/|\sin \gamma|) \exp [-\kappa D t] $ and
to recast it into a complete Fourier integral (denoted by the
tilde). This gives: 
\beq 
\kappa {\tilde \Phi}_{el} =  \int _{0} ^{u_1} du u^{-1} ( e^{-u} -1 ) \exp \left [- i
\tilde{q} \ln (u/u_0) \right ] 
\eeq 
with $u_m=u(m)$ and $\tilde q = \kappa ^{-1} {\bf q} \cdot {\bf \hat{v} }$. 
The complex solution of the integral reads 
\beq
\kappa {\tilde \Phi}_{el} = - u_0 ^{i\tilde{q}}
\left (  \frac{i u_{1}^{-i\tilde{q}}}{\tilde{q}} +\Gamma
(-i\tilde{q},u_1) -\Gamma (-i \tilde{q})        \right ) 
\eeq
in terms
of the complete and incomplete gamma functions, $\Gamma(z)$ and
$\Gamma(a,z)$ respectively. Because of the cosine transform in Eq. (\ref{phiel0}) we
only need the real part ${\text{Re}} (\kappa {\tilde \Phi}_{el})$. In
the asymptotic limit,
$u_1=A/|\sin \gamma| \gg 1 $ and the incomplete gamma function becomes
negligibly small so that it can be omitted in the remainder of the analysis.
Writing out the complex functions one arrives at 
\beq  
\kappa {\hat
{\Phi}}_{el} \approx \frac{\sin \tilde{q} \kappa D}{\tilde{q}}   + {
\text{Re}} \left [ u_0^{-i \tilde{q}} \Gamma (i \tilde{q}) \right ]
\label{complex} 
\eeq 
To further approximate the latter term  let us
first note that for the smectic and columnar modes we have
\begin{eqnarray} \tilde{q}_{S} & = & (\kappa L)^{-1} Q_{S} ({\bf
\hat{q}} \cdot {\bf \hat{v}}) \ll 1 \nonumber \\ \tilde{q}_{C} & = &
(\kappa D)^{-1} Q_{C} ({\bf \hat{q}} \cdot {\bf \hat{v}}) \sim
{\cal O}(1) 
\end{eqnarray} recalling that $\kappa D \sim {\cal O }(1)$  and $L/D
\gg 1$. For the smectic
symmetry, we may approximate  
\beq 
-\Gamma(i\tilde{q}) \approx
\gamma_{E} +i \tilde{q}^{-1}, \hspace{1cm} \tilde{q} \ll 1
\label{gamma} 
\eeq 
with $\gamma_{E}$  Euler's constant  to obtain:
\begin{eqnarray}
{\hat {\Phi}}_{el} &=& D j_{0} (\kappa D \tilde{q}_{S}) \nonumber \\
&& - \ik \left \{  \frac{ \sin \left ( \kappa \overline{D} \tilde{q}_{S}  \right )
}{ \tilde{q}_{S}}      -   \gamma_{E}  \cos \left ( \kappa
  \overline{D} \tilde{q}_{S}   \right )  \right \}
\label{chargesmec} 
\end{eqnarray}	 
with $ \overline{D} = D\left \{ 1 + (\kappa
D)^{-1} ( \ln A - \ln |\sin \gamma| )    \right \} $.  It is  easy
to verify that the electrostatic contribution vanishes in the hard rod
limit ($\kappa D \rightarrow \infty$) as it should.  Taking the limit
$\tilde{q}_{S} \rightarrow 0$ Eq. (\ref{chargesmec}) simplifies to:
\beq 
{\hat {\Phi}}_{el} = - \ik \left ( \ln A  - \ln |\sin \gamma | + \gamma_{E} \right ) \label{phiel}
\eeq	 
We may now compose the total Mayer kernel  by
replacing the thickness contribution  $j_{0}(D \bq \cdot {\bf \hat{v}})$ in Eq. (\ref{phitot}) by Eq. (\ref{phiel}). The contributions
depending on $L$ are left untouched since the line charge model, by definition, 
can only affect the interaction thickness of the rod. These considerations
lead to the following form of the Mayer kernel of two line charges: 
\beq 
{\hat \Phi}(\bq; \Omega _1,
\Omega _2  ) =  -  2 L^{2} D_{\text{eff}}(\Omega_1 , \Omega _2) | \sin
\gamma |  j_{0} ^{2} ( Q_{S}/2 ) \label{2vir}
\eeq 
valid for the smectic symmetry. It is similar to the first term in Eq. (\ref{smecor}) but with $D$
replaced by an orientation-dependent {\it effective thickness}
$D_{\text{eff}}$: 
\beq 
D_{\text{eff}}(\Omega _1, \Omega _2) = D \left
\{ 1 + \frac{1}{\kappa D} \left (\ln A - \ln |\sin \gamma | + \gamma _{E} \right ) \right \} 
\eeq 
which is exactly the same as the effective thickness showing up in the second virial
coefficient of two charged rods  in Ref. \onlinecite{stroobantslading}.

The next step is to perform the orientational average using
Eq. (\ref{0ODF}). The variational parameter $\alpha$ now follows from
minimizing the nematic free energy for {\it charged} rods. The
associated minimum condition reads \cite{stroobantslading}
\beq 
\alpha - {\cal K}_{1} c \alpha ^{1/2} + {\cal K}_{2} c \alpha^{1/2} (2-
\ln \alpha) = 0 \label{alfa} 
\eeq 
with constants ${\cal K}_{1} = \pi ^{-1/2}
[ 2+ (\kappa D)^{-1}(2\ln A + 3 \gamma _{E} -2) ]$ and ${\cal K}_{2} = \pi
^{-1/2}(\kappa D)^{-1} $. This equation has to be solved numerically
for any given concentration $c$ and electrostatic parameters $A$ and
$\kappa D$. One may verify the hard rod limit  $\kappa D \rightarrow
\infty$ to obtain $\alpha _{H} = 4c^{2}/\pi$.  The orientation average
of Eq. (\ref{2vir}) can be worked with the aid  of  Eq. (\ref{avsin})
and the following asymptotic result \cite{stroobantslading} 
\beq
\langle \langle \ | \sin  \gamma | \ln  | \sin \gamma | \rangle
\rangle _{f_{G}}\sim \left ( \frac{\pi}{4 \alpha} \right )^{1/2} (\ln \alpha
-2 + \gamma_{E}) 
\eeq 
It is expedient to normalize
the variational parameter  $\tilde{\alpha} = \alpha/\alpha _{H}$, so
that $\tilde{\alpha}$ approaches unity in the hard rod limit. Using
this, the nematic-smectic bifurcation condition Eq. (\ref{bif}) for
charged rods finally  becomes 
\begin{eqnarray}
1&=& -  4  j_{0} ^{2} ( Q_{S}/2)
\tilde{\alpha}^{-1/2} \nonumber \\ 
&& \times \left \{ 1 + \frac{1}{\kappa D} \left ( \ln A +
 \frac{1}{2} \ln \alpha + \frac{3}{2} \gamma _{E} - 1 \right ) \right
\}  \nonumber \\ 
&& - 8 \eta  j_{0}(Q_{S}) + \rho \hat{\Phi}_{Y}(Q_{S},0) 
\end{eqnarray}
with $\hat{\Phi}_{Y}$  given by Eq. (\ref{yuk}). Comparing this with
the hard rod result  Eq. (\ref{smfinal}) we see that the first
contribution is now implicitly dependent on concentration, albeit
weakly, due to the nonlinear character of Eq. (\ref{alfa}).

For the columnar symmetry we have to retain the complex gamma
functions in Eq. (\ref{complex}) and the resulting expression reads: 
\begin{eqnarray}
{\hat {\Phi}}_{el} &=&  D  j_{0} (\kappa D \tilde{q}_{C} )  +  \kappa ^{-1}  \{
\cos \left ( \kappa \overline{D} \tilde{q}_{C} \right ) {\text{Re}}
\left [  \Gamma \left ( i \tilde{q}_{C} \right ) \right ]  \nonumber \\
&& +   \sin
\left ( \kappa \overline{D} \tilde{q}_{C} \right ) {\text{Im}} \left [
\Gamma \left ( i \tilde{q}_{C} \right ) \right ] \}
\end{eqnarray}
With the aid of Eq. (\ref{gamma}), it is easily  shown that the
contribution vanishes in the limit $\kappa D \rightarrow \infty $
(corresponding to $\tilde{q}_{C} \rightarrow 0$).

We may combine this with the remaining contributions that depend only on the rod
length $L$. The  Mayer kernel of ${\cal O}(L^{2}D)$ for two line charges within the columnar symmetry can
then be written in the following form: 
\begin{eqnarray}
{\hat \Phi} &=& -2L^{2}D |\sin \gamma | j_{0} \left ( \frac{L}{2} \bq
\cdot  {\bf  \hat{w}_{1} } \right )  j_{0} \left ( \frac{L}{2} \bq
\cdot { \bf \hat{w }_{2} } \right ) \nonumber \\ 
&& \times G(\bq; \Omega_1, \Omega_2)
\label{collld} 
\end{eqnarray}
where $G$ now replaces the Bessel function
$j_{0}(D\bq \cdot {\bf \hat{v}})$ in Eq. (\ref{phitot}) 
\begin{eqnarray}
G(\bq;
\Omega_1, \Omega _2)  &=&  -( \kappa D)^{-1}  \{  \cos \left (
\kappa \overline{D}  \tilde{q}_{C}  \right ) {\text{Re}} \left [
\Gamma \left ( i  \tilde{q}_{C}  \right ) \right ] \nonumber \\ 
&&  +   \sin \left (
\kappa \overline{D} \tilde{q}_{C}  \right ) {\text{Im}} \left [
\Gamma \left ( i  \tilde{q}_{C}  \right ) \right ]  \} 
\end{eqnarray}
and the reader may verify that this contribution is recovered from $G$ in
the hard rod limit, as it should. With Eq. (\ref{yuk}) the {\it total}
Mayer kernel for the columnar symmetry is proposed to be of the following form:
\beq 
{\hat \Phi} = \hat{\Phi}_{L^{2}D}( \bq ; \Omega_1, \Omega_2 ) +
\hat{\Phi}_{LD^{2}}(\bq ; \Omega_1, \Omega_2 ) + \hat{\Phi}_{Y}(0, Q_{c})
\eeq 
with  $\hat{\Phi} _{L^{2}D}$ given by Eq. (\ref{collld}) and  $\hat{\Phi}
_{LD^{2}}$ the hard-core contribution of ${\cal O}(LD^2)$ from
Eq. (\ref{phitot}), both  depending intricately on the rod
orientations. Since the latter contribution  no longer pertains to the excluded volume of parallel
rods, as was the case for the smectic symmetry,  the
leading order term from the Yukawa site model $\hp_{Y}$ is expected to be 
orientation dependent too. However, getting access to this contribution
 requires  a full numerical integration over all spatial and
orientational variables of the HCY site model for freely rotating rod
pairs. Resolving the Mayer kernel would then give a 9 dimensional
integration (including the site-site summation) which clearly is a
formidable numerical task. Therefore, we shall rely on the parallel
contribution. The approximation can be justified in part from the
fact that the variational parameter $\alpha$ and hence the degree of
nematic order becomes rather large for sufficiently large Debye
lengths, as we will see later on. Finally, the nematic-columnar bifurcations can  be computed
by inserting the full Mayer kernel into the bifurcation condition
Eq. (\ref{colbif}) and performing a numerical averaging over the
orientational degrees of freedom using the Gaussian ODF.

\begin{figure*}
\includegraphics[width=\dfigwidth]{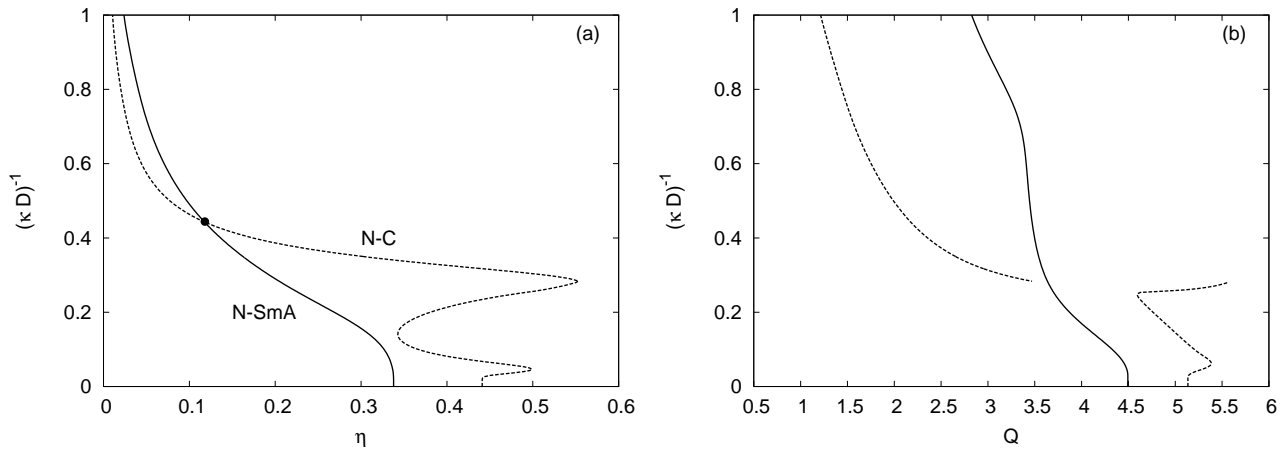}
\caption{(a) Bifurcation diagram for parallel charged rods. Plotted is the
normalized Debye length $\kappa^{-1}/D$ versus the hardcore volume
fraction $\eta$.  At $(\kappa D)^{-1} > 0.444$ a
stable nematic-columnar bifurcation pre-empts the nematic-smectic one.
The volume fractions are obtained from Parsons rescaling.  (b)
Dimensionless wave number $Q$ [Eq. (\ref{dwn})] corresponding to the bifurcation lines
shown in (a).} 
\end{figure*}

\section{Brownian dynamics simulations} To supplement the theory, 
Brownian dynamics (BD) simulations have been carried out for a system of
{\it point} Yukawa-site rods. The
corresponding interrod potential is virtually identical to Eq. (\ref{hcyuk})
with omission of the hardcore contribution for $s_{ij} < 1$ 
 (achieved by taking the limit $D \rightarrow 0$). This means that
overlaps of the inner cylindrical cores are in principle allowed in
our simulations. However these configurations are rare because of the
significant energy penalty involved. Another
difference with the description in Sec. IV is that the rods are no
longer fixed in parallel configurations but  are allowed to {\it
rotate freely}.  Apart from the rod length and the Bjerrum length,
the relevant lengthscale of the point Yukawa model is the Debye
length. The ratio of the latter two can be expressed as \beq \kappa \lambda _{B}
= \sqrt { 4 \pi Z \tilde{\rho} (\lambda _{B}/L)^{3} } \eeq in terms of
the dimensionless rod concentration $\tilde{\rho} = NL^{3}/V $.

\begin{figure}
\includegraphics[width= \columnwidth]{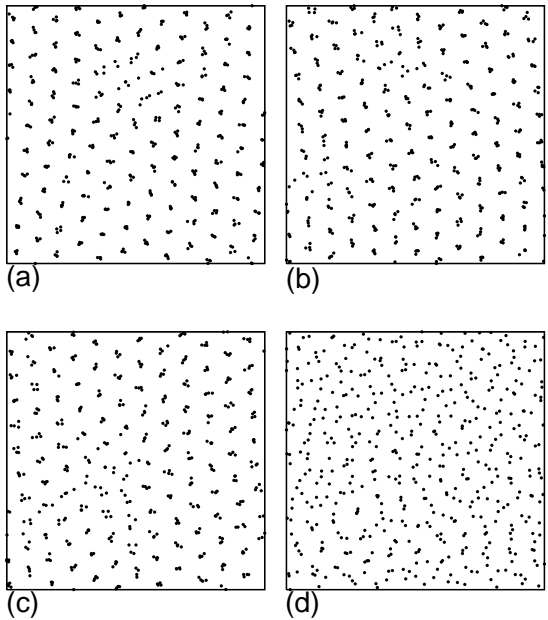}
\caption{Simulation snapshots showing the projections of the rod
centres-of mass (indicated by dots) onto the $xy$-plane perpendicular
to the field direction. Results correspond to various 
values for the external field strength $\xi$ [Eq. (\ref{uex})]. (a) parallel rods ($\xi
\rightarrow \infty$), (b) freely rotating rods at $\xi=50$, (c) same
for $\xi =25$ and (d) $\xi = 10$. }
\end{figure}
\begin{figure*}
\includegraphics[width=\dfigwidth]{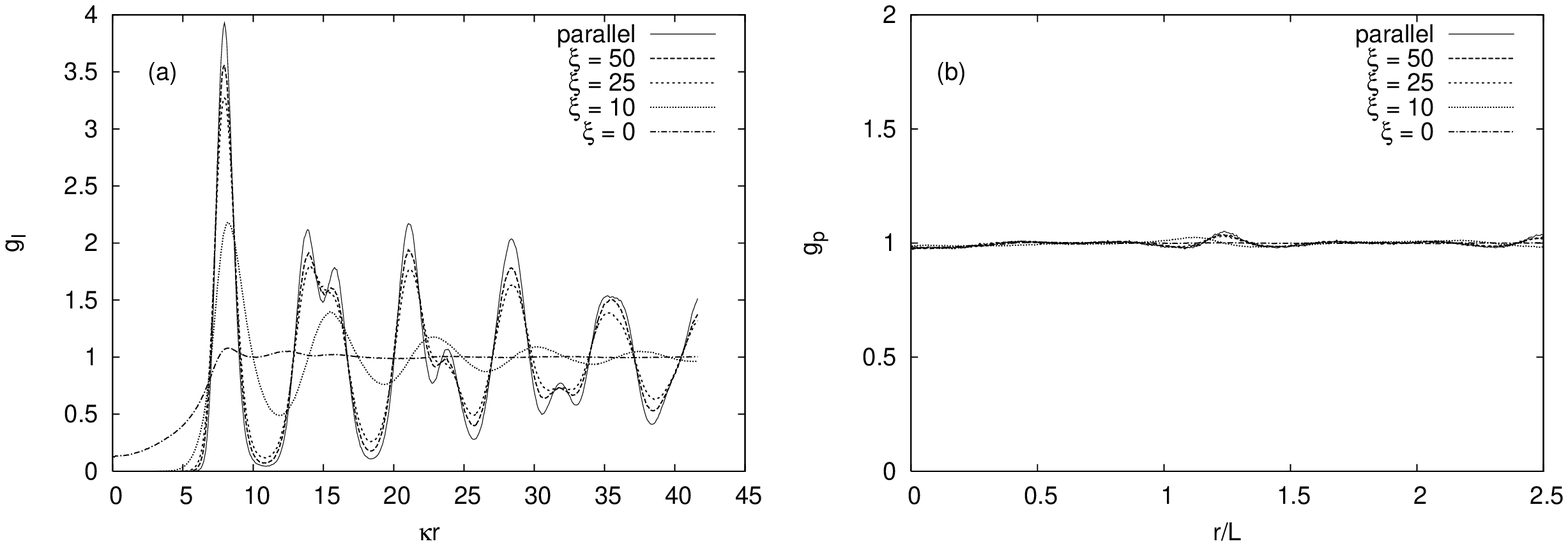}
\caption{(a) The intralayer and (b) the parallel pair correlation
functions for various field strengths $\xi$. }
\end{figure*}

The simulations comprise a finite difference
integration of the Langevin equations for interacting Brownian
macro-ions according to the scheme of Ermak \cite{ermak}. For a
detailed exposition of the update equations for Yukawa-site rods
the reader is referred to the paper of Kirchhoff {\em et al.}
\cite{kirch-bd}. The short-time self-diffusive behavior of the rods is
characterized by two translational diffusion coefficients (one
parallel and perpendicular to the rod axis) and a rotational one.
All of these depend on the hydrodynamic aspect ratio of the rods
\cite{tirado} for which we take a value of 16, comparable to that of
TMV rods \cite{fraden-tmv-baus}.

Systems consisting of $N=500$  rods each with $n=13$ sites were
simulated in a cubic box with periodic boundary conditions.  The
dimensionless concentration was fixed at a low value $\tilde{\rho} =
4.0$. Starting configurations were generated by randomly inserting
rods in a parallel, non-overlapping configuration. 
Equilibration is then carried out until the system has reached
a stable isotropic configuration, characterized by a vanishing nematic
order parameter, defined as the maximum eigenvalue of
the tensor \beq  {\bf Q} = \frac{1}{N}   \sum _{i} \frac{3}{2} \left
\langle  {\bf \hat{w}}_{i} \otimes {\bf \hat{w}}_{i} \right \rangle  -
\frac{1}{2}{\bf I}  \eeq where $\otimes$ denotes a dyadic product,
${\bf I}$ the second-rank unit tensor and the brackets a canonical
average.  The next step in our simulations is to {\it nematize} the
system by applying an external directional field along the
$z$-direction of the simulation box. This gives rise to an external
potential energy per rod, given by
 \beq \beta U_{i} = -\xi \cos ^2
({\bf \hat{w}}_{i} \cdot {\bf \hat{z}}) \label{uex}
\eeq 
in terms of a
dimensionless field strength $\xi >0$ (in units $k_{B}T$). Obviously, a stable nematic
state could also have been obtained by increasing the density of the
system. However, we found that the simulations become increasingly
cumbersome for large densities due to the slow dynamics and the
formation of metastable, transient states. Moreover, at high packing fractions
the results are  expected to be sensitive to the details of
the model (in particular the number of sites per rod) and many
time-expensive test runs need to be carried out. 

Once the directional
field is switched on, the (instantaneous) nematic order parameter is found to increase
rapidly until a plateau value is reached after some time interval. The
associated average nematic order parameter is found to be close to
unity for $\xi =50$ and $\xi =25$ (near parallel systems) and around
0.8 for the lowest nonzero applied field strength $\xi =10$. The
nematic state is then equilibrated further to allow for possible
translational freezing transitions.  To detect  smectic order 
(along ${\bf \hat{z}}$) or columnar order (perpendicular to ${\bf \hat{z}}$)
we first introduce the {\it intralayer} pair correlation function, defined
as \cite{cinacchi} 
\beq g_{l}(r) = \frac{1}{N \rho} \left \langle
\sum _{i} \sum _{j \neq i}  \delta (r - |\br _{ij} \times {\bf
\hat{z}}|) \Theta \left (L/2  - \br _{ij} \cdot {\bf \hat{z}}  \right
)  \right \rangle  
\eeq 
with $\delta$ the  Dirac delta function and
$\Theta$ the Heaviside step function.  It represents the probability
to find a particle at distance $r$ from a reference particle within a
slab of thickness $L$ perpendicular to the director. Basically,
$g_{l}$ provides  information about positional order perpendicular to
the nematic director and a profoundly peaked function is to be
expected in case of a columnar or crystal structure.
Second,  the {\it parallel} pair correlation function, given by \beq
g_{p}(r) = \frac{1}{N \rho} \left \langle  \sum _{i} \sum _{j \neq i}
\delta (r - \br _{ij} \cdot {\bf \hat{z}})  \right \rangle  \eeq
monitors liquid structure parallel to the nematic director and a
peaked $g_{p}$ will appear in case of smectic or crystalline  order.  To make sure the
results do not suffer from finite size effects, additional simulations
 have been carried out using $N=900$ rods with $n=13$ and no
qualitative differences were observed. Furthermore, the pressure
tensor \cite{ALLEN87} has been  monitored in all cases and no evidence
for spurious, non-isotropic stresses induced by the cubic  periodic boundary
conditions was encountered.

\section{Results and discussion}

Before discussing the results we first have to specify the
electrostatic parameters of the rod solution.  For these we take
typical values for TMV rods suspended in water ($\varepsilon _{r} =78
$) at neutral pH and room temperature. The rod dimensions are $D = 18$ nm
and $L=300$ nm and the Bjerrrum length is
$\lambda_{B}=0.716$ nm. The total rod charge is fixed at $Z=390$ and
the corresponding linear charge density is set at $\lambda_{B}\nu  =
1$ (i.e. one unit charge per Bjerrum length) which are reasonable
values if the effect of counterion condensation is taken into account
\cite{fraden-tmv-baus,Vroege92}. These parameters justify the use of  the
Debye-H\"{u}ckel approximation to determine the  potential amplitude in
Eq. (\ref{aa}) \cite{philipwooding}.

In Fig. 1 results are presented for parallel charged rods with $n=30$
HCY sites.  Increasing the number of sites per rod did not lead to
significant changes in the bifurcation diagram.
At high ionic strengths [$(\kappa D)^{-1} \leq 0.1$] the rod
charges are highly screened and the phase behavior is virtually  
unaffected by
the charge. A stable nematic-smectic bifurcation pre-empts the
nematic-columnar one.  At lower ionic strength the transition shifts to lower
volume fraction due to fact that the enhanced pair interactions 
lead to an effective `interaction' volume fraction higher
than the bare one. A qualitative change of scenario occurs  at even
lower ionic strengths  [$(\kappa D)^{-1} \geq 0.44$] where a 
 nematic-columnar instability pre-empts the nematic-smectic one and
a stable columnar phase can be expected. In this regime, the extension
of the double layers is significant and very little packing is
needed for the system to order.   The generic stabilisation of
 positional order is in accordance with the results of Ref. \onlinecite{kramerherzfeld-pre}
and \onlinecite{graftmv}. The enhanced
interaction range also shows up in the bifurcation wave numbers in
Fig. 1b where the sharp decrease implies a columnar spacing spanning multiple  rod
diameters. The  jump at $(\kappa D)^{-1}
\approx 0.28$ is a consequence of the fact that there are two independent bifurcating solutions
(i.e. a high-$Q$ branch and a low-$Q$ one) whose densities merge at that point.

The stability of the columnar state is confirmed by the simulation results
for the parallel Yukawa site model. The snapshot in Fig. 2a 
displays a hexagonal pattern indicative of freezing perpendicular to
the nematic director.  
The columnar nature of the structure is reflected explicitly in the pair
correlation functions shown in Fig. 3. As to $g_{l}$, the height of the first peak and the
double-peaked shape of the second  are a  hallmark of long-ranged
positional order.  More importantly, the flatness of $g_{p}$ signals an
 absence of long-range structure along the director (the small peak is due
 to weak  {\em intra}-columnar correlations). This means that there is
 no evidence of additional layering in the $z$-direction indicative of three-dimensional
crystalline order. 
 To roughly estimate whether
the state point adopted in the simulations corresponds to the
bifurcation diagram in Fig. 1  we may take the hydrodynamic aspect
ratio  $x_{h}=16.7$ and thickness to estimate $(\kappa D)^{-1} \sim
2.4 $ and $\eta \sim (\pi/4) \tilde{\rho} x_{h}^2 = 0.011$. Although the
screening length is beyond the scale depicted in Fig. 1 it can be
easily  inferred by extrapolation that the state point falls roughly in the
columnar stability regime.

Let us now turn to the case of freely rotating
rods. The main question is whether or not the scenario sketched above
is altered qualitatively if rotational degrees of freedom are allowed
for. 
\begin{figure}
\includegraphics[width=\figwidth]{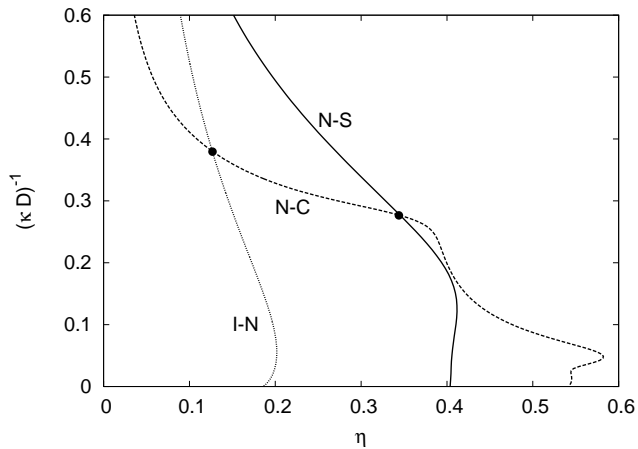}
\caption{ Bifurcation diagram for freely rotating charged rods. Plotted is the
normalized Debye length $\kappa^{-1}/D$ versus the hardcore volume
fraction $\eta$.  The isotropic-nematic (I-N) bifurcation is
obtained from Eq. (\ref{bifin}) in Appendix A.}
\end{figure}
The corresponding  bifurcation diagram  in Fig. 4 shows that
this is not the case. In fact, the pre-emption of the nematic-columnar
seems more outspoken here: the crossover point has shifted to
lower values and the the difference between the N-S and N-C volume
fraction is enhanced beyond the intersection point.
 However, in order to be able to make a final assessment of the stability of the
inhomogeneous phases  we have to verify the stability of the nematic
state itself. To illustrate this, the bifurcation from the isotropic
to the nematic state has been included in Fig. 4. This curve is
 obtained from a simple analysis described in Appendix A. 
Below the I-N bifurcation, the nematic phase is no longer stable
with respect to the isotropic and all
instabilities of the nematic state  located in this
region become
meaningless. For this reason, the scope of the bifurcation diagram is
bounded by the  intersection point at $(\kappa D)^{-1} \approx 0.4$.
Below this point, the overall scenario is a nematic-smectic transition
occurring at high screening $(\kappa D)^{-1} < 0.27$ while a gradual crossover
towards columnar order is expected at low salt conditions. The regime of low
screening [i.e. $(\kappa D)^{-1} > 0.4$], cannot be accessed  but it is
reasonable to anticipate translational symmetry-breaking  instabilities (of either
columnar or crystalline signature) of the
isotropic phase. These are beyond the scope of
the present calculations. The Gaussian variational parameter at the
instability is recorded in Fig. 5 and its large
value (the minimum is at $\alpha  \approx50 $) supports the use of the
asymptotic analysis adopted throughout this paper.

If the isotropic state were to be suppressed by applying a strong aligning 
external field, the nematic phase will be stable irrespective of
density and pronounced columnar order can be expected in a large portion of the
phase diagram, as we see in Fig. 4 (neglecting the I-N curve).
 This  is illustrated by the simulation results. Judging from the
 snapshots in Fig. 2b and 2c 
it is clear that the columnar structures are robust against small orientational fluctuations.
  The differences between the layer pair correlation
functions in Fig. 3  are very small. The number of lattice defects seems to increase slightly
upon lowering the field strength but  only at the smallest value
($\xi =10$) do we observe that the  hexagonal pattern has vanished completely. The system
nevertheless displays significant liquid like order perpendicular to the
director  which is  typical for  a dense nematic state. In the absence of
the field, the systems  is completely isotropic, in qualitative accordance with the
phase diagram in Fig. 4.

So far, we have implicitly assumed that the bifurcations represent
thermodynamically stable phase transitions. Although the simulations
clearly point to a (mechanically) stable columnar phase for both
parallel and asymptotically rotating rods, it would be desirable to
get similar confirmation from the theory. To that end we have
conducted a parametric expansion of the free energy around the
bifurcation, elaborated in Appendix B. The resulting Landau coefficient is depicted in Fig. 6. and it is clear
that the thermodynamic stability condition $C_{4}<0$ is generally
fulfilled  (except in the smectic region where N-S is
pre-empted by N-C). A similar outcome is found for
the parallel case, not shown here.

\begin{figure}
\includegraphics[width=\figwidth]{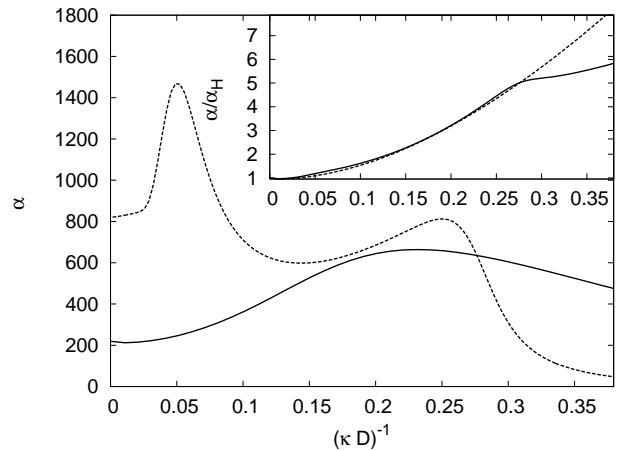}
\caption{ Behavior of the Gaussian variational parameter $\alpha$
  [see Eq. (\ref{alfa})] at the N-S
(solid) and N-C (dotted) bifurcations  as a function of Debye
length. The inset shows the normalized
$\tilde{\alpha}=\alpha/\alpha_{H}$ at the N-C bifurcation. }
\end{figure}
\begin{figure}
\includegraphics[width=\figwidth]{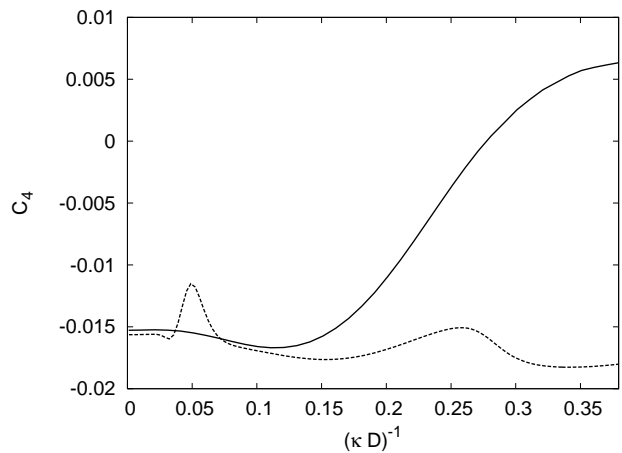}
\caption{Landau coefficient $C_{4}$ given by Eq. (\ref{c4}) at the
N-S (solid) and N-C (dotted) bifurcations  as a function of Debye
length corresponding to Fig. 4.}
\end{figure}

\section{Conclusions}

An extensive bifurcation analysis is presented on translational
symmetry-breaking instabilities in nematic systems of charged rods within
Onsager's second virial theory. Starting from an artificial
system of parallel charged rods, modelled via a hard-core Yukawa site
model, nematic-smectic and nematic-(hexagonal) columnar instabilities
are scrutinized as a function of the Debye length. The treatment is then extended towards the more realistic situation
of rods with rotational  degrees of freedom by employing an asymptotic
Gaussian analysis of the orientation-dependent quantities. 
In both cases, distinct preferential columnar order is observed for
moderate Debye screening lengths. As a supplement, Brownian dynamics simulations for
the point Yukawa site model were carried out and the stability of columnar
structures at low screening conditions  and low particle concentrations is
qualitatively reproduced.  No evidence for crystalline
order was found. Based on this, no attempt has been made in theory to
seek possible nematic-crystal instabilities, which could in principle be
scrutinized by imposing coupled smectic and columnar density
modulations. Moreover, these type of instabilities are more likely to
occur at very low screening conditions where the effective
rod aspect-ratio (which incorporates the extent of the double layers) is no longer
large enough to guarantee a stable nematic phase. In  this region one 
may
expect an isotropic fluid of rods to directly freeze into a crystalline lattice
without intervention of nematic order
\cite{mcgrother,Bolhuisintracing}. These type of bifurcations are not
addressed in this paper because the numerical effort involved in quantifying the necessary electrostatic end cap
contributions  for freely rotating rods is beyond the scale of the present calculations.

Looking at the experimental results for charged rods 
we may put forward  that a  crossover from smectic order
to a more intricate ordered state  has been observed in concentrated
systems of TMV rods  upon decreasing ionic strength \cite{fraden-tmv-baus}. However, at
present it is not fully clear whether these structures are really columnar or
represent three-dimensional crystalline order. More detailed structural
investigations are probably required to resolve this ambiguity. 

Our calculations also show that the columnar
order can  be realized at fairly low concentrations if the rods are
rendered in near-parallel configurations by an external
field. This could be any field that couples primarily to the rod
orientations, for example magnetic, electric or shear flow fields.
Finally, we remark that  our results are connected to observations
in other systems of rod-like mesogens with soft interactions. 
Columnar phases occur in complex systems of stiff polyelectrolytes like DNA \cite{DNAcol,livolantDNAoverview} and 
may be induced by an external magnetic field in systems of dipolar colloids \cite{holmweis} and 
lath-shaped, goethite colloids \cite{lemairecolumn}. Although
other complicating features such as length polydispersity, dipole-dipole or non-uniform site-site interactions
 are at play in these systems, it is intriguing to see that similar columnar structures may occur
in a relatively simple model system for soft rods considered here. Future work could be aimed at finding out whether a scenario such as in Fig. 1 
is qualitatively reproducible for
 other soft potentials. It is also desirable to enlarge the scope of the simulations 
such that a wide area of concentrations and rod potentials can be covered and better comparisons with theory and experiments can be made.

\section*{Appendix A: Isotropic-nematic bifurcation}

The bifurcation from the isotropic to the nematic state can be
analytically derived starting from the stationarity condition
Eq. (\ref{stat}) which we may rewrite as 
\beq \ln f(\Omega) = \mu +
\rho \int f(\Omega ^{\prime}) \hat{\Phi}(0; \Omega , \Omega ^{\prime}
)  d \Omega^{\prime} \label{statin} \eeq where the constant $\mu$
arises from the normalization condition of the ODF. The Mayer kernel
at zero wavevector is related to the second virial coefficient of two
charged rods [cf. Eq. (\ref{2vir})]:   \beq {\hat \Phi}(0; \Omega ,
\Omega ^{\prime}  ) =  -  2 L^{2} D_{\text{eff}}(\Omega , \Omega
^{\prime}) | \sin \gamma (\Omega , \Omega  ^{\prime}) | 
 \eeq 
The contributions of ${\cal O}(LD^{2})$ do not depend on orientation
and hence drop out of the free energy minimization with
respect to the ODF leading to Eq. (\ref{statin}).  Close to the
bifurcation  point the ODF can be parametrized as: \beq f(\Omega) =
\frac{1}{4 \pi} \left [ 1 + \epsilon {\cal P}_{2} (\cos \theta) \right
] \eeq with ${\cal P}_{n}$ a Legendre polynomial and $\epsilon$ an
arbitrarily small order parameter quantifying a uniaxial nematic perturbation
to the isotropic ODF $f=1/4\pi$.  Likewise, the orientation-dependent
functions in the Mayer kernel can be expanded in terms of these
polynomials in the following way \cite{stroobantslading}: \beq F( \sin \gamma ) = c_{0} +
c_{2}{\cal P}_{2} (\cos \theta) {\cal P}_{2} (\cos \theta ^{\prime}) +
\cdots  \eeq with coefficients $c_{0} = \pi/4 $ and $ c_{2} = -5 \pi
/32 $ for $ F(x)=x $  and $c_{0}=(\pi/4)[\ln 2 - 1/2] $ and $
c_{2}=(-5 \pi /32)[\ln 2 - 5/4]$ for $F(x)=-x \ln x$.  Inserting these
into the stationarity condition Eq. (\ref{statin}), linearizing with
respect to $\epsilon$ and some rearranging  (using the orthogonality
condition  $\int _{-1}^{1} d (\cos \theta) {\cal P}_{n}(\cos \theta)
{\cal P}_{m}(\cos \theta) =2\delta_{nm}/(2n+1)$) readily leads to a
closed expression for the bifurcation concentration:  \beq c ^{\ast}=
\eta ^{\ast} \frac{L}{D} =\frac{4}{1+ (\kappa D)^{-1} (\ln A +\gamma
_{E} - \ln 2 - \frac{5}{4})} \label{bifin} \eeq with $A$ given by
Eq. (\ref{aa}).

\section*{Appendix B: Landau expansion around  the bifurcation point}

An expression for the free energy difference $\Delta F = F_{I} -
F_{N}$ between the inhomogeneous state and the
homogeneous nematic reference state can be obtained by inserting the
parametrization Eq. (\ref{rhofourier}) into the  functional
Eq. (\ref{free}).  After some rearranging one can show that the free
energy difference per particle  reads: \beq \frac{ \Delta  \beta F}{N}
=  \frac{1}{2 \pi} \int _{0}^{2\pi} d \zeta W( \zeta) \ln W (\zeta) -
\frac{\rho}{4}  \sum _{l=1} ^{\infty} a_{l}^{2} \hat{\Phi}_{f} (l q)
\label{dfree} \eeq using the short-hand notation  $ \hat{\Phi}_{f} (l
q) = \langle \langle \hat{\Phi} (l \bq; \Omega, \Omega ^{\prime})
\rangle \rangle _{f} $ and \beq W( \zeta) = 1 +  \sum _{l=1} ^{\infty}
a_{l} \cos (l\zeta)  \eeq For the sake of brevity we will restrict
ourselves to the smectic case here, i.e. a single instability
mode. The analysis for the columnar symmetry  can be carried out in a
similar manner. Minimizing with respect to the order parameters
$a_{l}$ and the wave number $q$ leads to the following stability
conditions
\begin{eqnarray} A(a_{l},q,\rho) &=& \frac{1}{2 \pi} \int _{0}^{2\pi}
d \zeta \cos (m \zeta ) \ln W ( \zeta) \nonumber  \\ 
&& - \frac{\rho}{2}  a_{m} \hat{\Phi}_{f} (m q) = 0, \hspace{0.5cm} m \geq 1 \nonumber \\
B(a_{l},q,\rho) &=& \rho \sum _{l=1} ^{\infty} a_{l}^{2}
\hat{\Phi}_{f}^{\prime} (l q) = 0 \label{station}
\end{eqnarray} with $ \hat{\Phi}^{\prime} = \partial \hat{\Phi}
/\partial q  $. Let us now propose the following expansions in terms
of an arbitrarily small parameter $ \epsilon $
\begin{eqnarray} W( \zeta ; \epsilon) & = & 1 + \epsilon a_{1} \cos
\zeta + \epsilon ^2 a_{2} \cos 2 \zeta  + \epsilon ^ 3 a _{3} \cos 3
\zeta + \cdots \nonumber \\ \rho ( \epsilon ) & = & \rho _{0} +
\epsilon \rho _{1} + \epsilon ^2 \rho_{2} + \cdots \nonumber \\
q(\epsilon) & = & q_{0} + \epsilon q_{1} + \epsilon ^2 q_{2} + \cdots
\label{parameter}
\end{eqnarray} The expansion of $W$ is justified close to
bifurcation point where the spatial inhomogeneity is supposed to be
weak.  The zeroth order solution $\epsilon = 0$ reproduces the
bifurcation condition $\rho_{0} \hat{\Phi}_{f}(q) =1 $ and hence
$\{a_{1}, \rho_{0}, q_{0} \} = \{1,\rho^{\ast},q^{\ast}\}$.  Inserting
the parametrization into Eq. (\ref{station}) allows us to expand the
set of stationarity conditions as follows:
\begin{eqnarray} A(\epsilon)  &=& \epsilon \left ( A_{m}^{(0)} +
\epsilon A_{m}^{(1)} +  \epsilon ^2 A_{m}^{(2)} + \cdots \right ) =0,
\hspace{0.5cm} m \geq 1	         \nonumber \\ B(\epsilon)  &=&
\epsilon ^ 2 \left ( B ^{(0)} + \epsilon B ^{(1)} +  \epsilon ^2 B
^{(2)} + \cdots \right ) =0       \label{ab}
\end{eqnarray} The calculation of the coefficients is straightforward
but tedious and we will only give the essential results in the
remainder of the Appendix.  Performing an order by order solution of
the above set of equations yields up to first order
\begin{eqnarray} \rho_{1} &=&  0 \nonumber \\  q_{1} &=&  0 \nonumber
\\ a_{2} &=& \frac{1}{4 (1- \rho _{0} \hat{\Phi}_{f}  (2 q) ) }
\end{eqnarray} and up to second order (dropping the caret and the
subscript $f$ for notational clarity)
\begin{eqnarray}	 \rho_{2} &=&  \frac{ \frac{1}{8} \hp ^{\prime
\prime } (q)  (1-2a_{2})+ \rho _{0}  a_{2}^{2}  \hp ^{\prime }(q) \hp
^{\prime }(2q)  } {\frac{1}{2} \hp (q)\hp ^{\prime \prime }(q) - ( \hp
^{ \prime }(q) )^2  } \nonumber  \\ q_{2} &=&  \frac{ \frac{1}{4} \hp
^{ \prime } (q)  (1-2a_{2})+  a_{2}^{2}  \hp ^{ \prime }(2q)    }{\rho
_{0} ( \hp (q))^{2} -\frac{1}{2}  \hp ^{ \prime \prime }(q)   }
\nonumber  \\ a_{3} &=& \frac{\frac{1}{2} a_{2} -
\frac{1}{12}}{\rho_{0} \hp (3q) - 1 }
\end{eqnarray} The free energy difference can now be expanded in an
analogous way using the parametrization Eq. (\ref{parameter}) in
Eq. (\ref{dfree}). The Landau free energy as a function of the order
parameter $\epsilon$ reads \beq \frac{ \Delta \beta F}{N} = \epsilon
^2 C_{2} +  \epsilon ^3 C_{3} + \epsilon ^4 C_{4} \label{lanfree} \eeq with
coefficients
\begin{eqnarray} C_2 &=& 0 \nonumber \\
                 C_3 &=& 0 \nonumber \\ 
                 C_{4} &=& \frac{1}{64}
\left ( \frac{\rho _{0} \hp  (2q)} {1 - \rho _{0} \hp  (2q) } -1
\right ) \label{c4}
\end{eqnarray} In agreement with Mulder's results \cite{mulderparns} we get a zero cubic
contribution in Eq. (\ref{lanfree}) indicating the transition towards the inhomogeneous
state to be of second order. The thermodynamic stability of the new state
however is guaranteed by the condition $C_{4} < 0$ which has to be
assessed numerically via the Mayer kernel.

\acknowledgments I am grateful to R. Blaak for useful discussions and a critical reading of the manuscript.
 This work was supported by the  Deutsche
Forschungsgemeinschaft under SFB-TR6  and the  Alexander von
Humboldt Foundation.

\end{document}